\documentstyle[12pt]{article}
\def\abstract#1{\vskip 7mm 
	\begin{center}{\large Abstract}\par \bigskip
		\begin{minipage}[c]{12cm}
			\small #1
		\end{minipage}
	\end{center}
}
\def\title#1{\begin{center}{\Large\bf #1}\end{center}}
\def\author#1{\vskip 5mm \begin{center}{#1}\end{center}}
\def\address#1{\begin{center}{\it #1}\end{center}}

\newcommand{\bfr}{\begin{flushright}}
\newcommand{\efr}{\end{flushright}}

\begin{document}

\vspace*{-2cm}
\bfr{}\efr\vspace{-9mm}
\bfr{}\efr
\vspace{1cm}

\title{On the vacuum in the Moyal quantization}
\author{Takao KOIKAWA\footnote{E-mail: koikawa@otsuma.ac.jp}}
\vspace{1cm}
\address{
  School of Social Information Studies,
         Otsuma Women's University,\\
         Tama 206-0035,Japan\\
}
\vspace{2.5cm}
\abstract{ 
We study the features of the vacuum of the harmonic oscillator in the Moyal
quantization. The vacuums with and without using the normal ordering look 
different. The vacuum without the normal ordering is shown to be
expressed using the Weyl ordering. The Weyl ordered vacuum is then compared
with the normal ordered vacuum, and the implication of the
difference between them is discussed.  
}

\newpage
\setcounter{page}{2}
Many people are interested in the noncommutative approach because it is
the geometrical quantization using the functions. One may hope to
construct the quantum theory of gravity using the feature. In the
definition of the Moyal bracket, there appears one
parameter\cite{Moy}. It plays the role of the 
spacing between lattice points, when we formulate the soliton equations
in terms of the zero curvature equation with the Moyal bracket\cite{Koi}. The
same parameter is the Planck's constant in the Moyal quantization which
may be interpreted as the ''spacing'' of the Energy levels. This may
allude the link between the completely integrable models and the quantum 
theory. 

Lately the noncommutative soliton solutions\cite{Gopa} are found by
making use of the correspondence of the Moyal 
bracket and the operator formalism where the normal ordering is adopted
to construct the states. However, one of the advantages of using the Moyal
quantization lies in the fact that we use the functions instead of the
operators. We should discuss the features of the Moyal quantization
without resort to the operator formalism. 

One of the best way to understand the features of the Moyal quantization
should be to apply it to the familiar problem such as the harmonic
oscillator, and learn lessons from it. In the Moyal quantization of the
harmonic oscillator, we do not introduce the 
normal ordering as in the operator formalism. We obtain 
the vacuum by using the star product. Then other states are constructed
algebraically without resort to the operator formalism. 
It is algebraic because, instead of solving the differential equation, 
the states are obtained algebraically by making use of the
Dirac's Hamiltonian factorization\cite{Curt,Fair}. The correspondence
between the $\star$ genstates\cite{Fair} 
and the solutions obtained by solving the differential equations is
direct. We can construct the states starting with the vacuum. 

On the other hand, we can also define the vacuum using the normal 
ordered product in analogy with the operator formalism. Consequently there
appear two vacuums which do not look like the same. In order to compare
two vacuums we first show that the functions in the case that does not
adopt the normal ordering is expressed by using the Weyl ordering, and
then reorder this to the normal ordered one to make a comparison. These
vacuums are explicitly shown without resort 
to the operator formalism. The difference between two vacuums are made
clear, which would help us to understand the Moyal quantization.

We first recapitulate the notation of the Moyal algebra. The star product for functions $f=f(q,p)$ and $g=g(q,p)$  is defined by
\begin{equation}
f\star g = \exp \Bigg[ \hbar
\Bigg(\frac{\partial~}{\partial q}\frac{\partial~}{\partial\tilde p}-\frac{
\partial~}{\partial p}\frac{\partial~}{\partial\tilde q}\Bigg)
\Bigg] f({\bf x}) g({\bf \tilde x}) \vert_{{\bf x} = {\bf\tilde x}},
\label{eq:star1}
\end{equation}
where ${{\bf x}=(q,p)}$ and ${{\bf \tilde x}=(\tilde q,\tilde p)}$ and they are set equal after the derivatives are taken. 

In the analysis of the harmonic oscillator, it is more convenient to use the variables $z$ and its complex conjugate $\bar z$ instead of the variables $q$ and $p$. They are defined by
\begin{eqnarray}
z=q+ip,\\
z=q-ip.
\end{eqnarray}
We list up some formulas of the star products of these variables:
\begin{eqnarray}
z \star f(z,\bar z)&=&(z+\hbar \frac{\partial}{\partial \bar z})f(z,\bar z),
\label{eq:zfrml1}\\
\bar z \star f(z,\bar z)&=&(\bar z-\hbar \frac{\partial}{\partial  z})f(z,\bar z),
\label{eq:zfrml2}\\
f(z,\bar z) \star z&=&(z-\hbar \frac{\partial}{\partial \bar z})f(z,\bar z),\label{eq:zfrml3}\\
f(z,\bar z) \star {\bar z}&=&({\bar z}+\hbar \frac{\partial}{\partial  z})f(z,\bar z).
\label{eq:zfrml4}
\end{eqnarray}
Throughout this paper, we denote the star product commutation relation (CR hereafter) of $f(z,\bar z)$ and $g(z,\bar z)$ as
\begin{equation}
[f(z,\bar z),g(z,\bar z)]= f(z,\bar z)\star g(z,\bar z)-g(z,\bar z)\star f(z,\bar z).
\end{equation}

By using the above formulas and the definition of CR, we obtain the CR  between $z$ and $\bar z$:
\begin{equation}
[z,\bar z]=2\hbar,
\label{eq:zcr}
\end{equation}
which permits the interpretation that $z$ plays the role of the
annihilation operator and $\bar z$ the creation operator. By normalizing
them we introduce the variables $a$ and $a^\dagger$ by 
\begin{eqnarray}
a&=&\frac{z}{\sqrt 2},\\
a^\dagger&=&\frac{\bar z}{\sqrt 2}.
\end{eqnarray}
Then we can rewrite the above star product CR as 
\begin{equation}
[a,a^\dagger]=\hbar.
\label{eq:crofa}
\end{equation}

It is easy to derive 
\begin{eqnarray}
[z, (\bar z z)^n]&=&{2\hbar n{\bar z}^{n-1} z^n},
\label{eq:degree1}\\
\left[\bar z, (\bar z z)^n\right]&=& -2\hbar n\bar z^n z^{n-1}.
\label{eq:degree2}
\end{eqnarray}
Here the value $n$ appearing in the coefficients on the RHS can be understood as the degree of $\bar z$ and $z$, respectively.

Let us consider the classical Hamiltonian of the  harmonic oscillator given by
\begin{equation}
H=\frac{p^2}{2m}+k\frac{q^2}{2},
\end{equation}
where we shall assume $m=k=1$ hereafter. 
This is rewritten by using the star product of 
the variables $z$ and $\bar z$ as
\begin{equation}
H=\frac{1}{2}(p^2+q^2)=\frac{1}{2}{\bar z}z=\frac{1}{2}({\bar z}\star z+\hbar).
\end{equation}
By quantizing the Hamiltonian by use of the Moyal method, the eigenfunction
$f_n(x)$ belonging to the Energy level 
$(n+\frac{1}{2})\hbar$(n=0,1,2$\cdots$) is obtained, and it is expressed 
by using the Laguerre polynomial $L_n(x)$ as
\begin{equation}
f_n(x)=e^{-\frac{2x}{\hbar}}L_n(2x),(n=0,1,2,\cdots),
\end{equation}
where $x=(q^2+p^2)/2$. The eigenfunction $f_0$ 
belonging to the lowest level reads in terms of the star product
\begin{equation}
f_0={e}^{-\frac{2x}{\hbar}}={e}^{-\frac{\bar z z}{\hbar}}.
\label{eq:vac1}
\end{equation} 
This deserves to be called the vacuum, because we can show 
that it is annihilated by $z$ which plays the role of the vacuum annihilation: 
\begin{eqnarray}
& &z \star e^{-\frac{\bar z z}{\hbar}}
\nonumber\\
&=&z \star \sum_{n=0}^{\infty}\frac{1}{n!}(-\frac{1}{\hbar})^n(\bar z z)^n
\nonumber\\
&=&\sum_{n=0}^{\infty}\frac{1}{n!}(-\frac{1}{\hbar})^n\bar z^n z^{n+1}-\sum_{n=1}^{\infty}\frac{1}{(n-1)!}(-\frac{1}{\hbar})^{n-1}\bar z^{n-1} z^n
\nonumber\\
&=&0.
\end{eqnarray}
We can similarly show that
\begin{equation}
e^{-\frac{\bar z z}{\hbar}}\star \bar z=0.
\end{equation}
The higher states are constructed by repeatedly operating $\bar z$ from left
and $z$ from right by the star product on the vacuum.

On the other hand, it has been well known that the one dimensional harmonic oscillator is described by the creation and annihilation operators. The vacuum is defined by
\begin{equation}
:e^{-\frac{{\hat a}^\dagger \hat a}{\hbar}}:,
\label{eq:vac2}
\end{equation}
where the double dots denote the normal ordering and the operators $\hat
a^\dagger$ and $\hat a$. They satisfy the operator CR:
\begin{equation}
[\hat a, \hat a^\dagger]_{opr}=\hat a\hat a^\dagger-\hat a^\dagger \hat a=\hbar,
\end{equation}
which shows the isomorphism of the operator CR to the star 
product CR given in (\ref{eq:crofa}).

As an analogy of the operator formalism, we can define the vacuum. We
first define the normal ordering of the star products. 
The correspondence to the operator formalism is straightforward in this
case. When all the $a^\dagger$ are set to the left of $a$, we call
it  normal ordered in the same way as the operator case. 
We denote the normal ordering of m $a^\dagger$ and n $a$ by using the same double dots as in the operator product case as
\begin{equation}
:(a^\dagger)^m a^n:=a^\dagger \star a^\dagger \star \cdots a^\dagger \star a \star a \star \cdots \star a.
\end{equation}
By making use of the isomorphism of the CR of the creation and
annihilation operators and the star product CR of $a$ and $a^\dagger$, 
we define the vacuum by
\begin{equation}
:e^{-\frac{a^\dagger \star a}{\hbar}}:
=1+(-\frac{1}{\hbar})a^\dagger \star a+\frac{1}{2!}(-\frac{1}{\hbar})^2 
a^\dagger \star a^\dagger \star a \star a+\cdots.
\end{equation}
Then we can show that
\begin{eqnarray}
& &a \star :e^{-\frac{a^\dagger \star a}{\hbar}}:
\nonumber \\
&=&\sum_{n=0}^{\infty} \frac{1}{n!}(-\frac{1}{\hbar})^n:(a^\dagger)^n a^{n+1}:-\sum_{n=1}^{\infty} \frac{1}{(n-1)!}(-\frac{1}{\hbar})^{n-1}:(a^\dagger)^{n-1} a^n:
\nonumber \\
&=&0,
\end{eqnarray}
where the second term in the middle equation appears as the result of
the star product CR when $a$ multiplied from left passes through n
$a^\dagger$. We can similarly show that  
\begin{equation}
:e^{-\frac{a^\dagger \star a}{\hbar}}:\star a^\dagger=0.
\end{equation}

Now we have obtained two kinds of the vacuums for the same Hamiltonian of the
harmonic oscillator, one is with the normal ordering (\ref{eq:vac1}) and
the other is without it (\ref{eq:vac2}). 
The former is rewritten as
\begin{equation}
e^{-\frac{\bar z \star z +\hbar}{\hbar}}=e^{-1}e^{-\frac{\bar z \star z}{\hbar}} \sim e^{-\frac{a^\dagger \star a}{2\hbar}},
\end{equation}
which should be compared with the vacuum $(\ref{eq:vac2})$ defined using the
normal ordering. They do not look like the same at first glance.  This
should be pursued  
and we try to clarify this point in the present paper. 
One might think that all we have to do is just to reorder the above
expression to the normal ordered one and compare it with the
normal ordered vacuum, but things are not such easy. There is a problem,
that is, the above expression is not fully expressed in terms of the star
products when it is expanded. This does not occur as far as we use the
operators. Despite the CR of the operators and the star product CR are
of the same form, as far as the star products appear in the exponent, 
the difference from the operator formalism becomes manifest. 
Therefore we should be careful in its algebraic treatment. 
The expansion of the vacuum is given by
\begin{equation}
e^{-\frac{\bar z \star z +\hbar}{\hbar}}=1+(-\frac{\bar z \star z +\hbar}{\hbar}) +\frac{1}{2!}(-\frac{\bar z \star z +\hbar}{\hbar})(-\frac{\bar z \star z +\hbar}{\hbar})+\cdots.
\end{equation}
Here we should note that
\begin{equation}
(-\frac{\bar z \star z +\hbar}{\hbar})(-\frac{\bar z \star z +\hbar}{\hbar})\neq
(-\frac{\bar z \star z +\hbar}{\hbar})\star(-\frac{\bar z \star z +\hbar}{\hbar}),
\end{equation}
which makes it difficult to express all the products in terms of the star products only. 

In order to make it possible to compare this with the normal ordered one
where each term is expressed in terms of the normal ordered products, we
first show that $(\bar z z)^n$ is expressed by using the Weyl star 
product of $\bar z$
and $z$ which is made of the sum of all the possible permutations of
products of n $\bar z$ and n $z$ divided by the number of the
permutations. 
We shall denote the Weyl ordered star product of m $\bar z$ and n $z$ as
$(\bar z^m z^n)_W$. By denoting the sum of the possible permutations of
the star products of m $\bar z$ and n $z$ by $\rm{Perm}(\bar z^m,
z^n)$, 
it is written as
\begin{equation}
(\bar z^m z^n)_W=\frac{m!n!}{(m+n)!} \rm{Perm}(\bar z^m,z^n).
\end{equation}
We show that 
\begin{equation}
(\bar z z)^n=(\bar z^n z^n)_W.
\end{equation}
It is easy to see that this holds for $n=1$ case:
\begin{equation}
\bar z z=(\bar z z)_W=\frac{1}{2}(\bar z \star z+z \star \bar z).
\end{equation}
In order to prove this in general, we make use of the CR of $(\bar z^n
z^n)_W$ with $z$ and $\bar z$ by which the degree of $z$ and $\bar z$
are measured, respectively, as we mentioned before in
(\ref{eq:degree1}) and (\ref{eq:degree2}).

We first show that
\begin{equation}
\frac{1}{2}\{z \star (\bar z^n z^{n})_W-{(\bar z^n z^n)_W} \star z \}=n\hbar(\bar z^{n-1} z^n)_W. 
\label{eq:weylf2}
\end{equation}
In order to prove this, we let z operating from left on $(\bar z^n
z^n)_W$ in the 1st term on the LHS pass through it. Every time it passes 
through $\bar z$ it leaves a term proportional to $2 \hbar$ as can been
seen from the CR (\ref{eq:zcr}). Note that there are n $\bar z$ in every term
of $(\bar z^n z^n)_W$. The number of produced terms is
$n\times(2n!)/(n!)^2=(2n!)/(n!(n-1)!)$. Therefore when $z$ passes
through all the terms, we are left with the sum equal to the 2nd term on
the LHS and the sum of the newly produced terms. The newly produced
terms are obtained by dropping $\bar z$ impartially whenever z passes
through $\bar z$, they constitute $(\bar z^{n-1} z^n)_W$, but
some terms are identical. The multiplicity is obtained by dividing the
number of the produced terms by the number of terms in  $(\bar
z^{n-1} z^n)_W$, which becomes $2n$. These steps are shown below:
\begin{eqnarray}
& &\frac{1}{2}\{z \star (\bar z^n z^{n})_W-{(\bar z^n z^n)_W} \star z \}
\nonumber \\
&=&\frac{1}{2}\frac{(n!)^2}{(2n)!}\{z \star \rm{Perm}(\bar z^n z^n)_W-{\rm{Perm}(\bar z^n z^n)_W} \star z \}
\nonumber \\
&=&\frac{1}{2}\frac{(n!)^2}{(2n)!}(2\hbar)(2n) \rm{Perm}(\bar z^{n-1} z^{n})_W
\nonumber \\
&=&\frac{(n!)^2}{(2n-1)!}\hbar \rm{Perm}(\bar z^{n-1} z^{n})_W
\nonumber \\
&=&n\hbar(\bar z^{n-1} z^n)_W. 
\end{eqnarray}
We can rewrite the above formula (\ref{eq:weylf2}) in the CR form. We
write it here together with other CR which is shown in the similar way:
\begin{eqnarray}
[z, (\bar z^n z^n)_W] &=& 2\hbar n (\bar z^{n-1} z^n)_W,\\
\left[\bar z, (\bar z^n z^n)_W \right] &=& -2\hbar n(\bar z^n z^{n-1})_W,
\end{eqnarray}
which show that the degrees of $z$ and $\bar z$ are both $n$ as was discussed
in Eqs.(\ref{eq:degree1}) and (\ref{eq:degree2}).

Let us next show that
\begin{equation}
\frac{1}{2}\{z \star (\bar z^n z^{n})_W+{(\bar z^n z^n)_W} \star z \}=(\bar z^n z^{n+1})_W.
\label{eq:weylf1}
\end{equation}
To estimate this, we pick up a term in $(\bar z^n z^n)_W$ and let $z$
move among $\bar z$ 
simultaneously from left and right. Since $z$ commutes with other $z$ in 
the term, we do not need to take them into account. 
In passing through $\bar z$,
every time $z$ hits $\bar z$ and passes through it, it creates a term
proportional to $2\hbar$ by using the CR (\ref{eq:zcr}). Depending on
the number of $\bar z$ 
which $z$ passes through, the coefficients of the new terms differ. We
say the i-th $(i=0,1,2,\cdots,N)$ position when there are i $\bar z$ to
the left of the position. Note that it is the (N-i)-th position counting from
right. When we let $z$ pass from left to the i-th position, the
produced term is proportional to i, while it is -(N-i) when $z$ reaches
there from right. In order to make an
impartial distribution of z, we repeat such a procedure (N+1)times
varying the position i from 0 to N. Then note that 
the positions with the small
numbers are passed over many times by $z$ coming from left, 
and a little from right. Taking these into account, the newly produced
terms are estimated to be proportional to
\begin{equation}
i\times(N-i)-(N-i)\times i=0.
\end{equation}
Thus they cancel each other, and in the remaining terms $z$ are evenly
distributed by construction and so it is proportional to $\rm{Perm}(\bar
z^n, z^{n+1})$. Since some of the terms are identical we have to
evaluate the multiplicity $M$ which is obtained as 
\begin{equation}
M=\frac{2(n+1)\frac{(2n)!}{(n!)^2}}{\frac{(2n+1)!}{(n+1)!n!}}=\frac{2(n+1)^2}{2n+1}.
\end{equation}
Thus the LHS of Eq.(\ref{eq:weylf1}) reads
\begin{eqnarray}
& &\frac{1}{2}\{z \star (\bar z^n z^{n})_W+{(\bar z^n z^n)_W} \star z \}
\nonumber \\
&=&\frac{1}{2}\frac{(n!)^2}{(2n)!}\{z \star \rm{Perm}(\bar z^n, z^{n})+\rm{Perm}(\bar z^n, z^n) \star z \}
\nonumber \\
&=&\frac{1}{2(n+1)}\frac{(n!)^2}{(2n)!}\{(n+1)z \star \rm{Perm}(\bar z^n, z^{n})+(n+1)\rm{Perm}(\bar z^n, z^n) \star z \}
\nonumber\\
&=&\frac{1}{2(n+1)}\frac{(n!)^2}{(2n)!} \times M \times \rm{Perm}(\bar z^n,
z^{n+1}))
\nonumber \\
&=&\frac{(n+1)!n!}{(2n+1)!}\times \rm{Perm}(\bar z^n, z^{n+1})
\nonumber \\
&=&(\bar z^n z^{n+1})_W.
\end{eqnarray}

By using above formulas (\ref{eq:weylf2}) and (\ref{eq:weylf1}), 
we eventually arrive at one of the most important formulas in the
present discussion.
Since
\begin{eqnarray}
& &z\star(\bar z^n z^n)_W
\nonumber \\
&=&\frac{1}{2}\{z \star (\bar z^n z^{n})_W+{(\bar z^n z^n)_W} \star z \}
\nonumber \\
& & + \frac{1}{2}\{z \star (\bar z^n z^{n})_W-{(\bar z^n z^n)_W}
 \star z \},
\end{eqnarray}
we obtain
\begin{equation}
z\star(\bar z^n z^n)_W=(\bar z^n z^{n+1})_W+n\hbar(\bar z^{n-1} z^n)_W.
\label{eq:zonw1}
\end{equation}
In the similar way, we can show that
\begin{eqnarray}
(\bar z^n z^n)_W\star z&=&(\bar z^n z^{n+1})_W-n\hbar(\bar z^{n-1} z^n)_W,
\label{eq:zonw2}\\
\bar z \star (\bar z^n z^n)_W &=& (\bar z^n z^{n+1})_W-n\hbar(\bar z^{n-1} z^n)_W,
\label{eq:zonw3}\\
(\bar z^n z^n)_W\star \bar z &=& (\bar z^n z^{n+1})_W+n\hbar(\bar z^{n-1} z^n)_W.
\label{eq:zonw4}
\end{eqnarray} 

Now that $(\bar z z)^n$ are expressed in terms of the star products of $\bar z$ and $z$ by the Weyl ordered form, so is the vacuum:
\begin{equation}
e^{-\frac{\bar z z}{\hbar}}=\sum_{n=1}^{\infty} (\frac{-1}{\hbar})^n \frac{1}{n!}(\bar z^n z^n)_W.
\end{equation}
We can make sure that this is is really the vacuum by using the above formulas:
\begin{eqnarray}
& &z \star e^{-\frac{\bar z z}{\hbar}}
\nonumber\\
&=&\sum_{n=0}^{\infty} (\frac{-1}{\hbar})^n \frac{1}{n!}z \star (\bar z^n z^n)_W
\nonumber\\
&=&\sum_{n=1}^{\infty} (\frac{-1}{\hbar})^n \frac{1}{n!}\big( (\bar z^n z^{n+1})_W+n\hbar(\bar z^{n-1} z^n)_W \big)
\nonumber\\
&=&0,
\end{eqnarray}
and
\begin{eqnarray}
& &e^{-\frac{\bar z z}{\hbar}}\star \bar z
\nonumber\\
&=&\sum_{n=0}^{\infty} (\frac{-1}{\hbar})^n \frac{1}{n!}(\bar z^n z^n)_W \star \bar z
\nonumber\\
&=&\sum_{n=1}^{\infty} (\frac{-1}{\hbar})^n \frac{1}{n!}\big((\bar z^{n+1} z^n)_W+n\hbar(\bar z^n z^{n-1})_W \big)
\nonumber\\
&=&0.
\end{eqnarray}
Here Eqs.(\ref{eq:zonw1}) and (\ref{eq:zonw4}) are used.

We have shown that there are two kinds of vacuums and they are 
both expressed in terms of the infinite series of the star products of
$z$ and $\bar z$. In order to find their relations,
we rewrite the Weyl ordered vacuum by using the normal ordering 
and then compare it with the normal ordered vacuum.

In the operator formalism, it is known that, by using the Wick's
theorem, one can reorder the Weyl product to the sum of the normal
ordered product times the propagators. Also in the star product case, the
similar theorem holds. We give the results for small n and try to find
some rules appearing there. $(\bar z^n z^n)_n$ for small n are
\begin{eqnarray}
(\bar z z)_W&=&:\bar z z:+\hbar,\\
(\bar z^2 z^2)_W&=&:\bar z^2 z^2:+4\hbar :\bar z z:+2\hbar^2,   \\
(\bar z^3 z^3)_W&=&:\bar z^3 z^3:+9\hbar :\bar z^2 z^2:+18 \hbar^2 :\bar z z:+6 \hbar^3, \\
(\bar z^4 z^4)_W&=&:\bar z^4 z^4:+16\hbar :\bar z^3 z^3:+72 \hbar^2:\bar z^2 z^2:+96\hbar^3:\bar z z:
\nonumber\\
& &+24\hbar^4,\\
(\bar z^5 z^5)_W&=&:\bar z^5 z^5:+25\hbar:\bar z^4 z^4:+200\hbar^2:\bar z^3 z^3:+600\hbar^3:\bar z^2 z^2:
\nonumber\\
& &+60\hbar^4:\bar z z:+120\hbar^5,\\
(\bar z^6 z^6)_W&=&:\bar z^5 z^5:+36\hbar:\bar z^5 z^5:+450\hbar^2:\bar z^4 z^4:+2000\hbar^3:\bar z^3 z^3:
\nonumber\\
& &+5400\hbar^4:\bar z^2 z^2:+4320\hbar^5:\bar z z:+720\hbar^6.
\end{eqnarray} 
We shall expand the Weyl ordered vacuum in terms of the power of
$1/\hbar$ by substituting these results into the vacuum expression:
\begin{eqnarray}
& &e^{-\frac{\bar z z}{\hbar}}
\nonumber\\
&=&\sum_{n=0}^{\infty} (\frac{-1}{\hbar})^n \frac{1}{n!}(\bar z^n z^n)_W 
\nonumber\\
&=&1+\big(-1+\frac{2}{2!}-\frac{6}{3!}+\frac{24}{4!}-\frac{120}{5!}+\frac{720}{6!}-\cdots \big)
\nonumber\\
& &+(-\frac{1}{\hbar}) \big(-1+\frac{4}{2!}-\frac{18}{3!}+\frac{96}{4!}-\frac{600}{5!}+\frac{4320}{6!}-\cdots \big):\bar z z:
\nonumber\\
& &+(-\frac{1}{\hbar})^2 \big(\frac{1}{2!}-\frac{9}{3!}+\frac{72}{4!}-\frac{600}{5!}+\frac{5400}{6!}-\cdots \big):\bar z^2 z^2:
\nonumber\\
& &+O((\frac{1}{\hbar})^3)
\nonumber\\
&=&1+\big(-1+1-1+1-1+1-\cdots \big)
\nonumber \\
& &+(-\frac{1}{\hbar})\big(1-2+3-4+5-6+\cdots\big):\bar z z:
\nonumber\\
& &+(-\frac{1}{\hbar})^2 \frac{1}{2} \big(1-3+6-10+15-\cdots \big):\bar z^2 z^2:
\nonumber \\
& &+O((\frac{1}{\hbar})^3).
\end{eqnarray}
This shows that, in the coefficients of each order of
$O((\frac{1}{\hbar})^n)$, the alternating series appear and they are
not convergent. 
In order to find some relations with the normal ordered vacuum of which
the coefficient is convergent, we shall study the alternating 
series in detail. For the purpose of evaluating these series,
we truncate the infinite series at $(\frac{1}{\hbar})^N$ and calculate
the finite sum. 
They are given by
\begin{eqnarray}
O(1)&;&-1+1\cdots+ (-1)^N = \left\{
\begin{array}{@{\,}ll}
	-1, & \mbox{for odd N}\\
	0, & \mbox{for even N}
\end{array}
\right. \\
O(\frac{1}{\hbar})&;&1-2+3-\cdots+ (-1)^N N = \left\{
\begin{array}{@{\,}ll}
	\frac{N+2}{2}, & \mbox{for odd N}\\
	-\frac{N}{2}, & \mbox{for even N}
\end{array}
\right.\\
O((\frac{1}{\hbar})^2)&;&1-3+6-10+15+\cdots+(-1)^{N+1}\frac{N(N+1)}{2}
\nonumber \\
&=& \left\{
\begin{array}{@{\,}ll}
	(\frac{N+1}{2})^2, & \mbox{for odd N}\\
	-\frac{N(N+2)}{2}. & \mbox{for even N}
\end{array}
\right.
\end{eqnarray}  
If the limit of the infinite series should exist, they should be
obtained by taking $N \to \infty$. However as is seen from
above, they are oscillating and divergent. We also note that the way of
divergence is different for the $N$=even case and $N$=odd case.
At this point we might conclude bluntly that the normal ordered vacuum
and the Weyl ordered vacuum are different. But if we are generous
enough,  we note an intriguing feature from the above expansion. 
If we should get rid of the oscillating part at $O(1)$ and the
divergent part at $O(\frac{1}{\hbar})$ and $O((\frac{1}{\hbar})^2)$
and take $N \to \infty$ limit while keeping N odd, the remaining term becomes
\begin{eqnarray}
e^{-\frac{\bar z z}{\hbar}} &\sim& 1+(-\frac{1}{\hbar})\frac{1}{2}:\bar z z:+\frac{1}{2!}(-\frac{1}{\hbar})^2 \frac{1}{4} :\bar z^2 z^2:+O(\frac{1}{\hbar^3})
\nonumber \\
&=&1+(-\frac{1}{\hbar}):a^\dagger a:+\frac{1}{2!}(-\frac{1}{\hbar})^2:{a^\dagger}^2 a^2:+O(\frac{1}{\hbar^3}).
\end{eqnarray}
This suggests that the regularized Weyl ordered vacuum which we shall denote as $(e^{-\frac{\bar z z}{\hbar}})_{\rm{reg}}$ reads
\begin{equation}
 (e^{-\frac{\bar z z}{\hbar}})_{\rm{reg}}=:e^{-\frac{a^\dagger \star a}{\hbar}}:.
\end{equation}
Although we adopted a little bit artificial assumption such that $N$
should be odd in taking $N \to \infty$, 
the result seems reasonable from the physical viewpoint concerning 
the normal ordered product in the operator
formalism, which is introduced to get rid of the infinities. Also in the 
star product case we encounter such infinities in the Weyl ordered
vacuum, and the finite terms coincide with the normal ordered vacuum
up to $O((\frac{1}{\hbar})^3)$.

In summary, we showed two kinds of vacuums of the harmonic oscillator in the
Moyal quantization with and without the normal ordering. These are
studied within the star product framework without resort to the operator
formalism.  The direct counterparts to the Laguerre polynomial solutions 
are the states which is based on the vacuum with the Weyl
ordering. However when we study the vacuum in detail, it is divergent, 
which may be evaded by adopting the normal ordered vacuum instead. But
it would be done in compensation for the loss of clear correspondence between
the functions and $\star$ genstates.

\end{document}